# A doubly associated reference perturbation theory for water

Bennett D. Marshall

*ExxonMobil Research and Engineering, 22777 Springwoods Village Parkway, Spring TX 77389 USA*

## Abstract

In this work we develop a new classical perturbation theory for water which incorporates the transition to tetrahedral symmetry in both the dispersion and hydrogen bonding contributions to the free energy. This transition is calculated self-consistently using Wertheim's thermodynamic perturbation theory. However, since the reference fluid structure to the hydrogen bonding theory itself depends on hydrogen bonding, the theory represents an approach which goes beyond perturbation theory. The theory is shown to accurately represent the thermodynamics of pure water. It is demonstrated that the new theory can reproduce the anomalous density maximum as well as minima in the isothermal compressibility and isobaric heat capacity.

Bennettd1980@gmail.com

# I: Introduction

Equations of state (EoS) based on perturbation theory are widely used to predict the phase behavior and thermodynamic properties of complex mixtures. It is common to use Barker-Henderson second order perturbation theory (BH2)[1,2] to account for non-polar dispersion attractions. The statistical associating fluid theory (SAFT)[3–7] class of EoS employ Wertheim's thermodynamic perturbation theory (TPT)[8–10] to account for both hydrogen bonding and chain connectivity contributions to the free energy. Both BH2 and TPT assume that the structure of the fluid is unchanged by hydrogen bonding and remains that of the reference fluid. For water this assumption is violated. Water undergoes a structural transition to tetrahedral symmetry at ambient conditions[11–13] which results in the anomalous density maximum, isothermal compressibility minimum and isobaric heat capacity minimum. These features are completely missed in standard perturbation theories for water.

Recently Marshall[14–16] introduced the concept of associated reference perturbation theory (APT). Instead of the typical hard sphere reference fluid used in most perturbation theories, APT incorporates hydrogen bonding in the reference fluid to the BH2 perturbation theory for dispersion interactions. APT captures the transition to tetrahedral symmetry within the BH2 reference fluid and was demonstrated to give a substantial improvement in accuracy over the standard perturbation approaches. APT qualitatively reproduces the anomalous minima in both isothermal compressibility and isobaric heat capacity. However, APT was still not able to reproduce the density maximum observed for water.

In APT the transition to tetrahedral symmetry is set by the hydrogen bonded structure calculated through TPT. While the BH2 reference exhibits the appropriate transition to tetrahedral symmetry, the TPT contribution retains its hard sphere reference fluid. TPT assumes an unchanging reference fluid with the degree of association.

In this work we seek to build on APT and incorporate an associating reference fluid to TPT. To accomplish this, we incorporate the APT reference system integral within the TPT theory itself. The hydrogen bonding contribution is then reminimized subject to this new reference fluid, which allows for the self-consistent calculation of the resulting hydrogen bonded structure. We call the new approach doubly associated reference perturbation theory (DAPT). We then apply the approach to predict the phase behavior of water. DAPT and APT are shown to give similar predictions for many thermodynamic properties of water; however, unlike APT, DAPT is able to robustly reproduce the density maximum of water.

## II: Theory

In this section, we extend the associated reference perturbation theory (APT) of Marshall[14,15] to include the transition to tetrahedral symmetry in the hydrogen bonding reference fluid. This requires a complete rederivation of the hydrogen bonding theory. First, the course grained potential used to model water is described.

### a) The molecular model

Water is taken to be a single sphere of diameter $d$ with 4 hydrogen bonding sites: two hydrogen bond acceptor sites ($O_1$, $O_2$) and two hydrogen bond donor sites ($H_1$, $H_2$) in the overall set $\Gamma = \{O_1, O_2, H_1, H_2\}$. The overall potential of interaction between two water molecules is given as

$$\varphi(12) = \varphi_{hs}(r_{12}) + \varphi_{as}(12) + \varphi_{sw}(r_{12}) \tag{1}$$

Where (1) represents the position $\mathbf{r_1}$ and orientation $\mathbf{\Omega_1}$ of molecule 1, $\varphi_{hs}$ is the potential of the hard sphere reference fluid and $\varphi_{sw}$ is an attractive perturbation due to an isotropic square well attraction of depth $\varepsilon$ and range $\lambda$. The association contribution to the pair potential is evaluated using a sum over site-site contributions[8]

$$\varphi_{as}(12) = \sum_{A \in \Gamma} \sum_{B \in \Gamma} \varphi_{AB}(12) \qquad (2)$$

The site-site contributions $\varphi_{AB}$ are approximated with conical square well (CSW) association sites[17]

$$\varphi_{AB}(12) = \begin{cases} -\varepsilon_{AB} & r_{12} \leq r_c \text{ and } \theta_{A1} \leq \theta_c \text{ and } \theta_{B2} \leq \theta_c \\ 0 & \text{otherwise} \end{cases} \qquad (3)$$

Where $r_c$ is the maximum distance between molecules for which association can occur, $\theta_{A1}$ is the angle between the center of site $A$ on molecule 1 and the vector connecting the two centers, and $\theta_c$ is the maximum angle for which association can occur. While CSW association sites give a very simplistic view of hydrogen bonding, they incorporate the short ranged and highly directional nature of the hydrogen bond, which leads to a limited valence. From a mathematical standpoint, CSW association sites are convenient because they allow for a decoupling of orientational and radial degrees of freedom.

### b) The Helmholtz free energy

A relatively simple, general and accurate approach to develop equations of state from model potentials such as Eq. (1) is perturbation theory. Perturbation theory assumes the structure of the fluid is dominated by short ranged forces. The longer ranged attractive contributions to the free energy can then be added as a perturbation to the reference (short ranged) fluid structure. A typical application of perturbation theory to a fluid of associating hard spheres with square well attractions would yield the following Helmholtz free energy

$$a = \frac{A}{Nk_bT} = a_{ref} + a_{sw} + a_{as} \tag{4}$$

In Eq. (4), $N$ is the number of molecules in the fluid and $T$ is the absolute temperature. The free energy contribution $a_{ref}$ gives the free energy of the reference fluid (typically taken as a hard sphere reference $a_{hs}$) , $a_{sw}$ the excess free energy due to isotropic square well attractions, and finally $a_{as}$ is the excess free energy due to association.

In this work we evaluate $a_{sw}$ with a simple first order perturbation theory[1]

$$a_{sw} = -2\pi \frac{\varepsilon}{k_bT} \rho d^3 I_{ref}(\lambda) \tag{5}$$

where $\rho$ is the number density of molecules and $I_{ref}(\lambda)$ is the integral of the reference system correlation function over the square well range $\lambda$

$$I_{ref}(\lambda) = \int_1^{\lambda/d} x^2 g_{ref}(x) dx \qquad x = r/d \tag{6}$$

$I_{ref}(\lambda)$ is related to the coordination number within a shell of diameter $\lambda$ through the relation

$$N_{ref}(\lambda) = 4\pi\rho I_{ref}(\lambda) \tag{7}$$

Hence $I_{ref}(\lambda)$ controls the number of molecules which will participate in isotropic square well attractions with a molecule located at the origin. In our previous associated reference perturbation theory (APT)[14], second order Barker-Hendersen perturbation theory (BH2)[1] was employed for square well contribution; however, we have found that for water the additional complexity of the second order theory does not improve the methodology over a simple first order approach.

The reference system fluid is often taken as a hard sphere fluid ($I_{ref} = I_{hs}$). This approach will perform well if the fluid structure is not substantially changed by association. An example of where this is not the case is water. Liquid water undergoes a structural transition to tetrahedral symmetry as temperature is decreased. That is, the structure of the fluid is substantially changed by association. For this reason, perturbation theories typically perform poorly for the phase behavior of pure water.

Remsing *et al.*[18] demonstrated using molecular simulation for SPC/E water that accurate perturbation theories could be obtained for water if one includes the short-ranged attractive forces (association) in the definition of the reference fluid. Marshall embraced this idea and developed associated reference perturbation theory (APT).[14] In APT, hydrogen bonding contributions are included in the reference system to the BH2 perturbation theory. Equation (4) is rearranged as

$$a = a_{ref} + a_{sw}$$

$$a_{ref} = a_{hs} + a_{as}$$

(8)

Specifically, tetrahedral symmetry is enforced when all water molecules are fully hydrogen bonded, and the theory reduces to the standard hard sphere reference fluid in the absence of hydrogen bonding. The key derivation of APT is the new reference system integral

$$I_{ref}(\lambda) = I_{hs}(\lambda) + \chi_4 \left( \frac{1}{\pi \rho d^3} - I_{hs}(r_c) \right)$$

(9)

where $\chi_4$ is the fraction of water molecules bonded 4 times. All association contributions to the theory where calculated within Wertheim's first order perturbation theory (TPT1)[9].

APT was demonstrated to be vastly superior to hard sphere reference perturbation theories in the thermodynamic description of pure water. APT correctly exhibits minima in both the isothermal compressibility and isobaric heat capacity, while hard sphere perturbation theories (HSPT) do not. Also, APT was shown give a substantially improved representation of liquid phase densities as compared to HSPT. However, while APT did give a substantial improvement, the thermodynamic description of water was not completely satisfactory. When applied to pure water, APT does not predict the density maximum.

In the original formulation of APT, the association contribution to the free energy $a_{as}$ was treated within TPT1[9,10] with a hard sphere reference fluid. That is, the associating reference fluid integral in Eq. (9) was not used in the calculation of the hydrogen bonding state of the fluid. In this work we include the transition to tetrahedral symmetry in the reference fluid for the hydrogen bonding free energy. Including these changes will require a complete rederivation in the hydrogen bonding contribution to the free energy. We develop the approach in Wertheim's multi-density statistical mechanics[8]. However, by definition, we are going beyond perturbation theory, as the reference fluid structure will be dependent on the hydrogen bonding state of the system.

### c) A rederivation of the association theory to enforce transition to tetrahedral symmetry

Within Wertheim's reformulation of statistical mechanics[8], each bonding state of a molecule is given its own density $\rho_\alpha$, where α is the set of bonded sites. Hence, $\rho_o$ is the monomer density of species $k$. To aid in the reduction from fugacity to density graphs, Wertheim defines a set of density parameters

$$\sigma_\alpha = \sum_{\gamma \subset \alpha} \rho_\alpha \tag{10}$$

Two special cases of Eq. (10) are for the monomer $\sigma_o = \rho_o$ and total densities $\sigma_\Gamma = \rho$, where Γ represents the set of all bonding sites. Also, the total density of molecules not bonded at the set of sites α is given by $\sigma_{\Gamma-\alpha}$.

The change in Helmholtz free energy due to association is given by[9]

$$\frac{A_{as}}{Vk_BT} = \rho \ln\left(\frac{\rho_o}{\rho}\right) + Q + \rho - \frac{\Delta c^{(o)}}{V} \tag{11}$$

with

$$Q = -\rho + \sum_{\substack{\gamma \subset \Gamma^{(k)} \\ \gamma \neq \emptyset}} c_\gamma \sigma_{\Gamma-\gamma} \qquad (12)$$

The functions $c_\gamma$ are generated from the graph sum $\Delta c^{(o)}$ according to the relation

$$c_\gamma = \frac{\partial}{\partial \sigma_{\Gamma-\gamma}} \frac{\Delta c^{(o)}}{V}; \quad \gamma \neq \emptyset \qquad (13)$$

The graph sum $\Delta c^{(o)}$ contains all the information on the hydrogen bonding interactions. In thermodynamic perturbation theory (TPT) only the contributions to $\Delta c^{(o)}$ which contain a single associated cluster are retained. This allows for the summation of $\Delta c^{(o)}$ in terms of reference system contributions with a single association bond are retained. All non-cyclic clusters can be constructed from this single irreducible diagram, but second order effects such as steric hindrance cannot be included.

The graph sum in TPT1 is given by[8,19]

$$\frac{\Delta c^{(o)}_{TPT1}}{V} = \frac{1}{2} \sum_{A \in \Gamma} \sum_{B \in \Gamma} 4\pi \kappa_{AB} d^3 f_{AB} I_{hs}(r_c) \sigma_{\Gamma-A} \sigma_{\Gamma-B} \qquad (14)$$

Where $\kappa_{AB}$ is the probability that site $A$ on molecule 1 is oriented correctly to bond to site $B$ on molecule 2

$$\kappa_{AB} = \frac{(1-\cos\theta_c)^2}{4} \qquad (15)$$

The hard sphere reference fluid integral $I_{hs}$ does not change with the degree of association. To include this information in the theory, we replace $I_{hs}$ in Eq. (14) with $I_{ref}(r_c)$ evaluated with Eq. (9). With this modification the fundamental graph sum is given by

$$\frac{\Delta c^{(o)}}{V} = \frac{1}{2} \sum_{A \in \Gamma} \sum_{B \in \Gamma} 4\pi \kappa_{AB} f_{AB} d^3 \sigma_{\Gamma-A} \sigma_{\Gamma-B} \left( I_{hs}(r_c)(1-\chi_4) + \frac{\chi_4}{\pi \rho d^3} \right) \qquad (16)$$

In the limit $\chi_4 \to 0$ the original TPT1 result is recovered. We approximate $\chi_4$ by assuming independent association sites[20]

$$\chi_4 = (1-X_{O_1})(1-X_{O_2})(1-X_{H_1})(1-X_{H_2}) \tag{17}$$

where $X_{O_1}$ is the fraction of water molecules not bonded at site $O_1$

$$X_{O_1} = \frac{\sigma_{\Gamma-O_1}}{\rho} \tag{18}$$

We will require the derivatives

$$\frac{\partial \chi_4}{\partial \sigma_{\Gamma-O_1}} = -\frac{(1-X_{O_2})(1-X_{H_1})(1-X_{H_2})}{\rho} \tag{19}$$

Assuming the fractions not bonded are the same for all sites we obtain for each site $A$

$$\chi_4 = (1-X_A)^4 \tag{20}$$

and

$$\frac{\partial \chi_4}{\partial \sigma_{\Gamma-A}} = -\frac{(1-X_A)^3}{\rho} \tag{21}$$

Employing Eqns. (13) and (16) and enforcing equivalence of sites

$$c_A = 8\pi \kappa_{AB} d^3 f_{AB} \left[ \rho X_A \left( I_{hs}(r_c)(1-\chi_4) + \frac{\chi_4}{\pi \rho d^3} \right) + 2\rho X_A^2 (1-X_A)^3 \left( I_{hs}(r_c) - \frac{1}{\pi \rho d^3} \right) \right] \tag{22}$$

$$c_\alpha = 0 \quad for \quad n(\alpha) > 1 \tag{23}$$

Equation (23) implies the following result for the unbonded fractions as well as the monomer fraction[21]

$$X_A = \frac{1}{1+c_A} \tag{24}$$

$$X_o = \frac{\rho_o}{\rho} = X_A^4 \tag{25}$$

Combining Eqns. (20), (22) and (24) provides a closed equation for the calculation of $X_A$. Eqns. (12), (24) and (25) are now combined to yield

$$\frac{Q}{\rho} + 1 = \sum_{\substack{\gamma \subset \Gamma^{(k)} \\ \gamma \neq \emptyset}} c_\gamma X_\gamma = 4c_A X_A = 4(1 - X_A) \tag{26}$$

Using Eqns. (25) and (26) the association free energy Eq. (11) is simplified to

$$\frac{A_{as}}{Nk_B T} = 4(\ln X_A + 1 - X_A) - \frac{\Delta c^{(o)}}{N} \tag{27}$$

where $\Delta c^{(o)}$ is simplified from Eq. (16)

$$\frac{\Delta c^{(o)}}{N} = 16\pi f_{AB} \kappa_{AB} d^3 \rho X_A^2 \left( I_{hs}(r_c)(1 - \chi_4) + \frac{\chi_4}{\pi \rho d^3} \right) \tag{28}$$

To evaluate the association contribution to the chemical potential we employ the exact relation from Wertheim[8]

$$\frac{\mu_{as}}{k_b T} = \ln X_o - c_o = 4\ln X_A - c_o \tag{29}$$

where $c_o$ is evaluated by taking the total density derivative of $\Delta c^{(o)}/V$ while holding all other densities $\sigma_\alpha$ constant

$$c_o = 16\pi \kappa_{AB} f_{AB} d^3 \rho^2 X_A^2 \left( \frac{\partial I_{hs}(r_c)}{\partial \rho}(1 - \chi_4) + \left( \frac{1}{\pi \rho d^3} - I_{hs}(r_c) \right) \frac{\partial \chi_4}{\partial \rho} - \frac{\chi_4}{\pi \rho^2 d^3} \right) \tag{30}$$

with

$$\frac{\partial \chi_4}{\partial \rho} = 4(1 - X_A)^3 \frac{X_A}{\rho} \tag{31}$$

This completes the development of the association theory which accounts for the transition to tetrahedral symmetry. This associated reference provides the reference fluid for the isotropic square well contribution through Eq. 9. Since the associated reference itself has an associated reference, we will refer to this approach as the doubly associated reference perturbation theory (DAPT). However, the association theory developed in this section is not a perturbation theory. TPT assumes the structure of the reference fluid remains unchanged by association. Clearly this is not the case here.

Figure 1 compares DAPT calculations for the fraction of unbonded sites $X_A$, association contribution to the chemical potential $\mu_{as}$, and association contribution to the compressibility factor $Z_{as}$ to TPT1 predictions. We assume the same potential parameters used in the original APT approach of Marshall[14] ($\varepsilon_{AB} = 1891.8\ k_bT$, $r_c = 1.294d$, $\theta_c = 28°$) and a constant liquid like reduced density of $\rho^* = \rho d^3 = 0.7$.

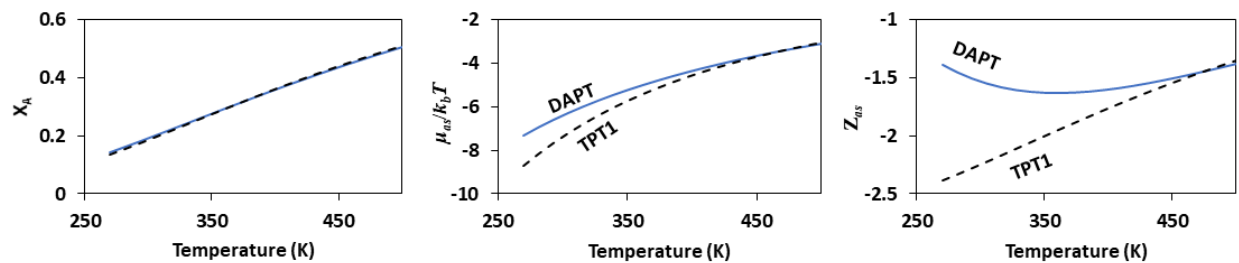

**Figure 1:** Comparison of DAPT predictions (solid curves) to TPT1 (dashed curves) for the fraction of unbonded sites (left), association contribution to the chemical potential (center), and association contribution to the compressibility factor (right). Density is held constant at $\rho^* = 0.7$

As can be seen, both TPT1 and DAPT give nearly identical predictions for the fraction of unbonded sites. However, the two theories give substantially different predictions for $\mu_{as}$ and $Z_{as}$. DAPT results a weaker decrease in the chemical potential due to association. The temperature dependence of $Z_{as}$ is remarkable. TPT1 shows a linear increase in $Z_{as}$ with increasing $T$, while DAPT exhibits a minimum near $T = 360\ K$. The difference between these two results is the inclusion of the transition to tetrahedral symmetry in the development of the association model within DAPT.

At this point we can be optimistic that the $Z_{as}$ minimum in DAPT may result in a density maximum in water which is absent in the APT / TPT1 theory. However, the parameters used in the DAPT calculations in Fig. 1 where from the TPT1 optimization. To validate the DAPT approach a new set of parameters will need to be developed for DAPT. This will be the subject of section III.

**III: Parameterization and model results**

In this section we develop model parameters for DAPT and compare model predictions to the thermodynamic properties of water. DAPT is described by the parameters $\varepsilon$, $\lambda$, $d$, $\varepsilon_{AB}$, $r_c$ and $\theta_c$. We fix $\theta_c = 30°$ and adjust the remaining parameters to reproduce vapor pressure $P_{sat}$ and saturated liquid density $\rho_L$ data in the temperature range 273 K $\leq T \leq$ 573 K. We found that the choice of $r_c$ set the temperature of the density maximum. Hence, a value of $r_c$ was chosen which gave a good description of the density maximum, with the remaining parameters $\varepsilon$, $\lambda$, $d$, $\varepsilon_{AB}$ chosen to minimize the error in the prediction of $P_{sat}$ and $\rho_L$ data over the full temperature range.

| $d$(Å) | $\varepsilon / k_b$ (K) | $\lambda / d$ | $\varepsilon_{AB} / k_b$ (K) | $\theta_c$ | $r_c / d$ |
|---|---|---|---|---|---|
| 2.931 | 254.23 | 1.900 | 1744.84 | 30° | 1.193 |

**Table 1:** Model parameters

| AAD% $P_{sat}$ | AAD% $\rho_L$ |
|---|---|
| 1.5 | 0.30 |

**Table 2:** Average absolute deviation between DAPT and experiment

Table 1 gives the model parameters and Table 2 list the average absolute deviations (AAD) between DAPT and experiment. The diameter $d$ is only slightly larger than measured through neutron diffraction data of Soper et al.[22], which shows that the first maximum in oxygen-oxygen correlation function in liquid water is located at a distance of $d = $ 2.75 Å. The critical radius $r_c = $ 3.5 Å is consistent with the first minima in the oxygen-oxygen correlation function as measured from this same neutron diffraction data[22]. From spectroscopy data, Luck[23] estimated the energy of a liquid phase water-water hydrogen bond to be $1862k_b$ which is similar to the regressed result in Table 1.

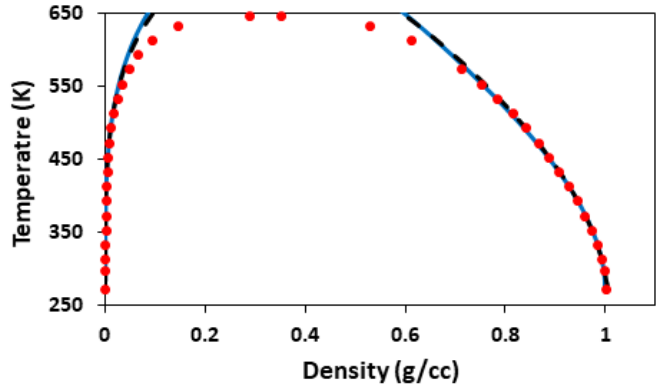

**Figure 2:** $T$-$\rho$ phase diagram for water. Solid curve DAPT model results, dashed curve gives APT model results and symbols give experiment[24]

In what follows we compare DAPT and the APT of Marshall[14] to experimental data for a variety of thermodynamic properties. Figure 2 gives results for the temperature-density phase diagram and Fig. 3 gives results for $P_{sat}$ and the heat of vaporization $H_{vap}$. Both DAPT and APT give very similar predictions of these quantities. Both theories are nearly visually indistinguishable.

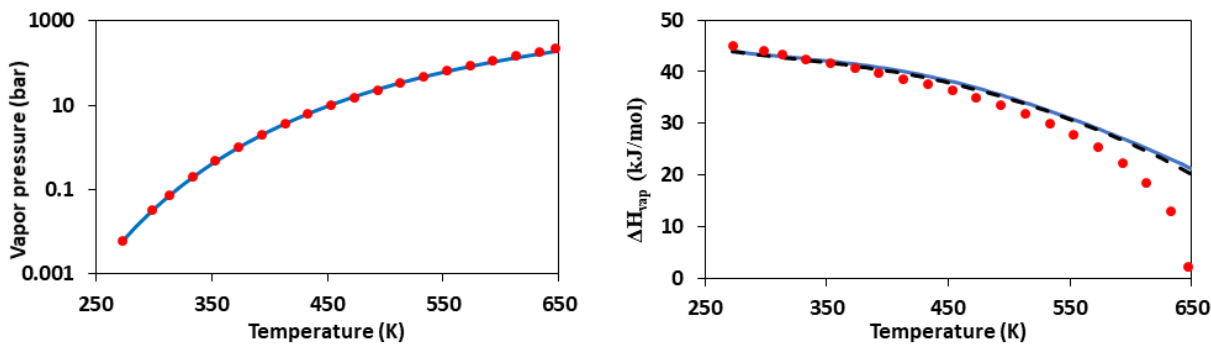

**Figure 3:** Comparison of model to experimental data[24] for $P_{sat}$ (left) and $H_{vap}$ (right). Symbols and curves have same meaning as Fig. 2

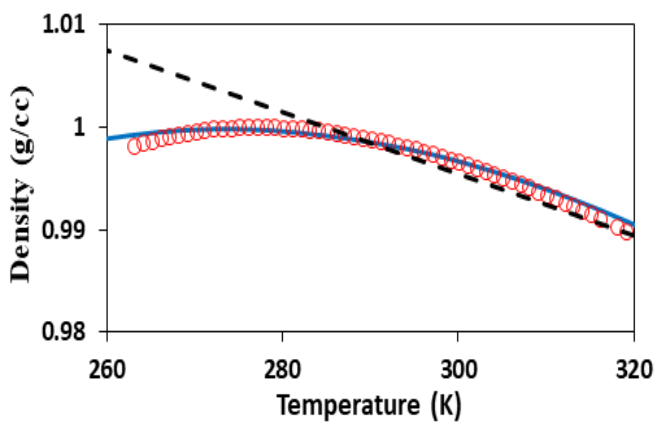

**Figure 4:** Liquid density[25] as a function of temperature at atmospheric pressure. Curves and symbols have the same meaning as Fig. 2

Both DAPT and APT give similar good agreement with the experimental data for the results in Figs. 2-3; However, DAPT accurately represents the density maximum of water, while APT does not give a density maximum. This can be seen in Fig. 4, which compares DAPT and APT predictions of liquid density at atmospheric pressure to experimental data.

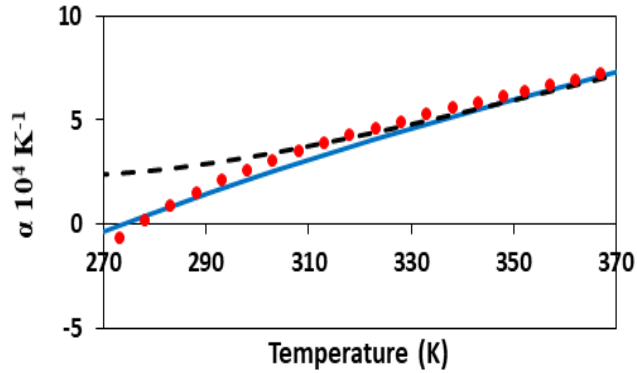

**Figure 5:** Volume expansivity as a function of temperature[26]. Curves and symbols have same meaning as Fig. 2

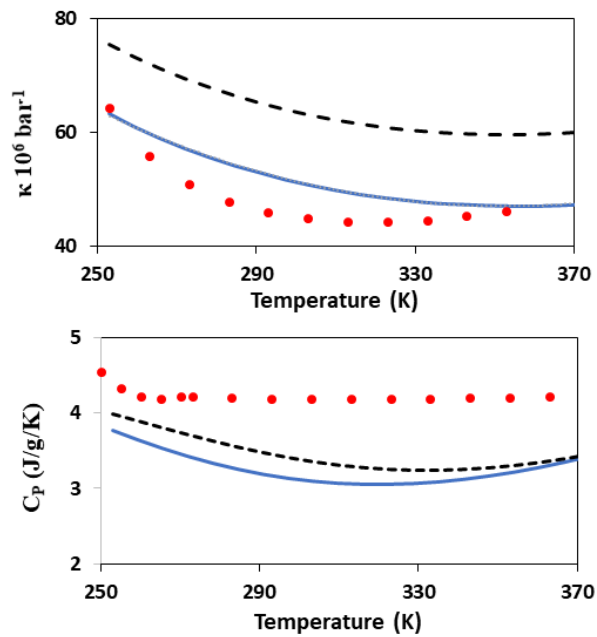

**Figure 6:** Isothermal compressibility[25] (top) and isobaric heat capacity[27] (bottom) at atmospheric pressure versus temperature. Lines and symbols have the same meaning as Fig. 2

This difference can be further observed in the volume expansivity α shown in Fig. 5. DAPT reproduces the anomalous α < 0 while APT does not exhibit this anomaly. The improved predictions of DAPT over APT are the result of including the transition to tetrahedral symmetry within the association contribution itself.

Figure 6 compares theory prediction to data for the isothermal compressibility and isobaric heat capacity at atmospheric pressure. As can be seen, both APT and DAPT exhibit the anomalous minima in these quantities. DAPT gives a substantial improvement over APT for the prediction of the isothermal compressibility; however, both approaches underpredict the heat capacity.

Lastly, Fig. 7 compares theory predictions for the fraction of free OH groups (assumed to be $X_A$) to the spectroscopic data of Luck[23]. Both APT and DAPT overpredict this fraction, which implies that both theories are predicting a lower degree of hydrogen bonding as compared to the results of Luck. The exact interpretation of Luck's data is difficult. Application of TPT to water has met with mixed results[7,15,16,23,28,29] in the description of Lucks data. It has been shown[16,28] that Luck's data is consistent with the results neutron diffraction results of Soper *et al.*[22] as well as molecular dynamics simulations using the *i*AMOEBA[29] polarizable classical force field. However, this consistency will depend on the strict definition of hydrogen bonding.

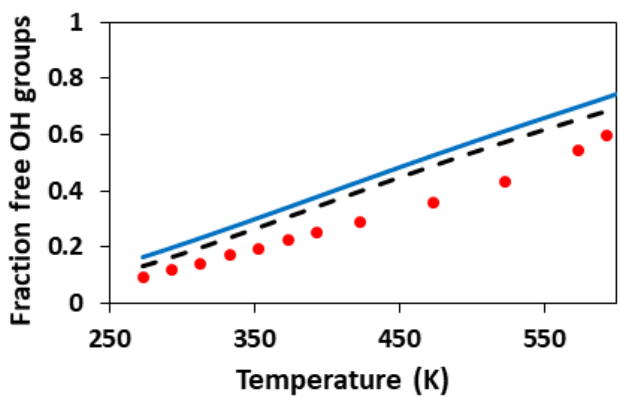

**Figure 7:** Comparison of theoretical predictions of the fraction of free OH groups to the data of Luck[23]. Lines and symbols have the same meaning as Fig. 2

IV: Conclusions

APT and DAPT give an accurate equation of state for water. Both approaches give a good overall representation of liquid density, vapor pressure, heat of vaporization and qualitatively reproduce the minima in isothermal compressibility and isobaric heat capacity. However, DAPT allows for the representation of the density maximum, while APT does not yield a density maximum. The difference between the two approaches is that DAPT includes the transition to tetrahedral symmetry in both the association and square well contributions, while APT only considers this transition in the square well term. The inclusion of this transition to tetrahedral symmetry in the association reference fluid results in non-monotonic behavior in the association contribution to the pressure as illustrated in Fig. 1. It is this non-monotonic behavior which allows for the representation of the density maximum.

Marshall[16] recently developed a resumed thermodynamic perturbation theory (RTPT) for water which incorporates hydrogen bond cooperativity. To demonstrate the accuracy of the DAPT approach in a clear and simple way, we chose to build on the standard TPT1 formalism instead of the RTPT for hydrogen bond cooperativity. However, any complete physical theory for water must grapple with the non-additive effects of hydrogen bond cooperativity. Future work will focus on the marriage of RTPT and DAPT to develop a complete theory for water which properly incorporates the transition to tetrahedral symmetry, as well as the non-additivity of hydrogen bonding.